\documentclass[journal,onecolumn]{IEEEtran}

\usepackage{amsmath}
\usepackage{amssymb, amsfonts}
\usepackage{amsthm}
\usepackage{epstopdf}
\usepackage{mathrsfs}
\usepackage{epstopdf}
\usepackage{hyperref}
\usepackage{graphicx, url, color}

\newcommand\floor[1]{\lfloor#1\rfloor}

\newtheorem{mydef}{Definition}
\newtheorem{theorem}{Theorem}
\newcommand{\C}{\mathbf{C}}
\newcommand{\D}{\mathbf{D}}
\newcommand{\La}{L}

\newcommand{\U}{\mathbf{U}}
\newcommand{\V}{V}
\newcommand{\W}{\mathbf{W}}
\newcommand{\X}{\mathbf{X}}
\newcommand{\Z}{\mathbf{Z}}
\newcommand{\G}{\mathbf{G}}

\newcommand{\Q}{\mathcal{Q}}
\newcommand{\Xin}{\mathbf{X}}
\newcommand{\Y}{\mathcal{Y}}

\newcommand{\mb}[1]{\mathbf{#1}}

\newcommand{\remove}[1]{}
\newcommand{\blu}[1]{{\color{black} #1}}
\newcommand{\bluer}[1]{{\color{black} #1}}

\begin{document}
\author{
   \IEEEauthorblockN{
   Vaishakh Ravindrakumar\IEEEauthorrefmark{1}, Parthasarathi Panda\IEEEauthorrefmark{2}, 
     Nikhil Karamchandani\IEEEauthorrefmark{3}, and
     Vinod Prabhakaran\IEEEauthorrefmark{4}
    }\\
\IEEEauthorblockA{\IEEEauthorrefmark{1}
     Dept. of Electrical and Computer Engg., UC San Diego, Email: vaishakhr@ucsd.edu}\\
\IEEEauthorblockA{\IEEEauthorrefmark{2}
     Amazon India, Email: parthasarathipanda314@gmail.com}\\
\IEEEauthorblockA{\IEEEauthorrefmark{3}
     Dept. of Electrical Engg., IIT Bombay, Email: nikhilk@ee.iitb.ac.in}\\
   \IEEEauthorblockA{\IEEEauthorrefmark{4}
     School of Technology and Computer Science, TIFR, 
     Email: vinodmp@tifr.res.in} 
 }

\title{Private Coded Caching}
\maketitle

\begin{abstract}
Recent work by Maddah-Ali and Niesen (2014) introduced \textit{coded caching} which demonstrated the benefits of joint design of storage and transmission policies in content delivery networks. They studied a setup where a server communicates with a set of users, each equipped with a local cache, over a shared error-free link and proposed an order-optimal caching and delivery scheme. In this paper, we introduce the problem of \textit{private coded caching} where we impose the additional constraint that a user should not be able to learn anything, from either the content stored in its cache or the server transmissions, about a file it did not request. We propose a feasible scheme for this setting and demonstrate its order-optimality by deriving  information-theoretic lower bounds.
\end{abstract}

\section{Introduction}
\label{Sec:Intro}
Broadband data consumption has grown at a rapid pace over last couple of decades, owing in great part to multimedia applications such as Video-on-Demand \cite{CiscoReport}. Content delivery networks attempt to mitigate this extra load on the communication network by deploying storage units or caches where some of the popular content can be pre-fetched during the off-peak hours. 

Content caching and delivery has been studied extensively in the literature, see for example~~\cite{Wessels:2001, breslau99, Borst:2010} and references therein. However, most of the work proposes caching schemes where those parts of the requested files that are available at nearby caches are served locally and the remaining parts are served by a remote server via separate unicast transmissions to the users. In contrast, recent work \cite{CachingUM, DCachingUM} has studied an information-theoretic formulation of the problem and proposed the idea of \textit{coded caching} which uses the available cache memory to not only provide local access to content but to also generate coded-multicasting opportunities among users with different demands. The setup studied in \cite{CachingUM, DCachingUM} consists of a server communicating to a set of users, each equipped with a cache of uniform size, over a broadcast link and the objective is to minimize the worst-case server transmission rate, over all feasible user demands. For this setup, coded caching is shown to provide significant benefits over traditional caching and delivery, and is in fact within a constant factor of the optimal. In this work, we consider a similar problem setup where we have the additional constraint that no user should be able to obtain any information, from its cache content as well as the server transmission, about any file other than the one it has requested. We call this setup `\textit{private}' and devise a \textit{private coded caching} scheme. \blu{Since we require that the cache content at each user not reveal any information about the files, but that along with the server transmission each user be able to reconstruct its demanded file, our formulation naturally suggests using the  idea of \textit{secret sharing} \cite{cramer2015}, \cite{shamir1979} as part of our solution strategy. Under secret sharing, a set of `shares' is generated for each file so that the file can be reconstructed if one has access to all the corresponding shares, but that any subset of the shares of size less than a threshold reveals nothing about the file.

We derive an achievability scheme based on secret sharing. Further, incorporating the condition of privacy on top of the cut-set based lower bound in~\cite{CachingUM}, we derive a lower bound on the server transmission rate. Comparing the achievable rate with the derived lower bound, we demonstrate that the performance of our scheme is within a constant factor of the optimal for several parameter values of interest. For instance, if the number of files is greater than the number of users, our scheme is order-optimal for all feasible values of cache memory. We refer to Theorem~\ref{Thm:OrderOptimality} for more details. Following the setting of \cite{DCachingUM}, we also propose a \textit{decentralized} variant of our achievability scheme,  wherein caching is carried out independently at each user. This mode of caching allows for varying number of users and for the absence of a centralized coordinating server. The performance of this \textit{decentralized} scheme is again shown to be within a constant factor of the optimal.}

The results of \cite{CachingUM, DCachingUM} on coded caching have been extended in several other directions as well, ranging from heterogeneous cache sizes \cite{wang2015fundamental}, unequal file sizes \cite{zhang2015coded2}, to improved converse arguments \cite{Ajaykrishnanetal15, ghasemi2015improved, sengupta2015improved}. Content caching and delivery has also been studied in the context of  device-to-device networks, multi-server topologies, and heterogeneous wireless networks in  \cite{ji2013wireless}, \cite{shariatpanahi2015multi}, and \cite{FemtoCaching} respectively. The work closest to ours is \cite{sengupta2015}, which considers the problem of \textit{secure coded caching}, where the goal is to protect information about the files from an eavesdropper which can listen to the server transmissions. \blu{However, \cite{sengupta2015} does not capture the notion of \textit{privacy} that we consider here. Throughout the paper, by \textit{security} / \textit{secure} we mean protection against an eavesdropper and by \textit{privacy} / \textit{private} we refer to our problem of interest. While most of the work in this paper focuses on private coded caching, we will also briefly discuss the case where one requires both privacy and security.}

The rest of the paper is organized as follows. We describe the problem setup in Section~\ref{Sec:Setup} and present our main results in Section~\ref{Sec:Results}. We discuss some examples and then describe our proposed \blu{\textit{centralized}} private coded caching scheme in Section~\ref{Sec:Scheme}. We present converse arguments and prove order-optimality of the proposed scheme in Section~\ref{Sec:Converse}. A \textit{decentralized} variant of the proposed scheme is presented in Section~\ref{Sec:Dec} and we conclude with a discussion of our results in Section~\ref{Sec:Discussion}. \blu{A part of this work was presented in \cite{scc2016}; this manuscript has the complete proofs of all the results as well as original technical content. In particular, Section~\ref{Sec:Dec} is new where we present a \textit{decentralized} private coded caching scheme where the content placement in the caches is carried out independently at each user. We further show order-optimality of the proposed decentralized scheme.}

\section{Problem Formulation} 
\label{Sec:Setup}
\noindent{\em Notation:} For $n \in \mathbb{N}$, we denote by $[n]$ the set $\{1,2,\ldots, n\}$. A vector of random variables will be denoted by bold-faced upper case letters, e.g., $\mb{Y}=(Y_1,Y_2,\ldots,Y_n)$. For a set $A\subseteq [n]$, we will denote the vector of random variables indexed by elements in $A$ by $\mb{Y}_A$. Specifically, if $A=\{i_1,i_2,\ldots,i_m\}$ where $1\leq i_1 < i_2 < \ldots < i_m \leq n$, we denote $\mb{Y}_A=(Y_{i_1},Y_{i_2},\ldots,Y_{i_m})$.

\begin{figure}
\centering
\includegraphics[scale=1]{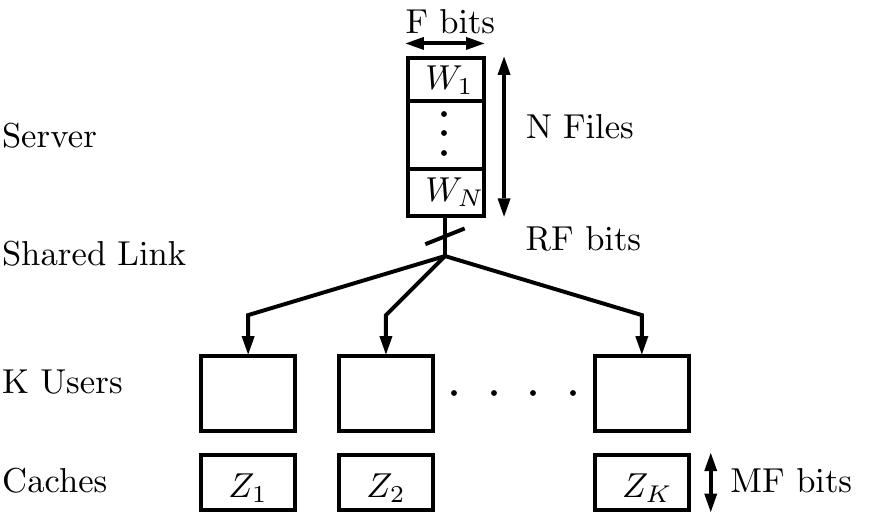}
\caption{In the setup above, a server containing $N$ files, each of $F$ bits, is connected via an error-free shared link to $K$ users, each with a cache memory of size $MF$ bits. The server multicasts through this link at a rate of at most $RF$ bits.}
\label{fig:setup}
\end{figure}
We consider a single-hop content delivery network, as illustrated in Figure~\ref{fig:setup}. The system consists of a server hosting a collection of $N$ files, $\mb{W} = (W_1, W_2,\ldots,W_N)$, each of size $F$ bits. We will assume that $W_1,W_2,\ldots,W_N$ are independent random variables each distributed uniformly over $[2^F]$. The server is connected via a shared, error-free link to $K$ users, each with a cache memory of size $MF$ bits. We will refer to $M$ as the normalized cache memory size. 

The system works in two phases: a \textit{placement phase} followed by a \textit{delivery phase}. In the placement phase, the user caches are populated with content related to the $N$ files using a possibly randomized scheme. Formally, we denote the content stored in cache $k$ by a random variable $Z_k$ which takes values in $[2^{MF}]$. The vector $\mb{Z} = (Z_1,Z_2,\ldots,Z_K)$ is jointly distributed according to some conditional distribution $p_{\mb{Z} |\mb{W}}$. Note that the placement phase is performed without any prior knowledge of future user demands. During the delivery phase, each user requests one of the $N$ files. The resulting demand vector $\mb{d} = (d_1,d_2,\ldots,d_K)$ is revealed to all the users and the server. The server transmits a message $X_{\mb{d}}(\mb{W}, \mb{Z})$ of size $RF$ bits on the shared link to the users. 

Each user $k$ generates an estimate $\widehat{W}_{d_k}$ of its requested file $W_{d_k}$ using only its stored cache content $Z_k$ and the server transmission $X_{\mb{d}}(\mb{W}, \mb{Z})$. The probability of error of a placement and delivery scheme is given by
\begin{align} 
P_e\triangleq\max_{\mb{d}\in[N]^K} \mathbb{P}\big((\widehat{W}_{d_1},\ldots,\widehat{W}_{d_K})  \neq (W_{d_1},\ldots,W_{d_K})\big), \label{err}
\end{align}
where the probability is over the files and the randomization in the placement phase, i.e., over the distribution $p_{\mb{W}}p_{\mb{Z}|\mb{W}}$. \blu{Note that we take the worse-case error probability over all possible demand vectors $\mb{d}$.} 
In addition to recovering the demanded files, we also want each user to not obtain any information about the other files. The information leakage of a placement and delivery scheme is defined as:
\begin{equation} \label{sec}
L\triangleq \max_{\mb{d} \in[N]^K}\max_{k\in [K]}I(\mb{W}_{[N] \setminus \{d_k\}}; X_{\mb{d}}(\mb{W}, \mb{Z}),Z_k).
\end{equation}
A placement and delivery scheme is said to be an $(\epsilon, \delta)$-\textit{private scheme} if its probability of error $P_{e} \le \epsilon$ and information leakage $L \le \delta$. 

\begin{mydef} \label{def:privately-achievable} The memory-rate pair $(M,R)$ is said to be {\em privately achievable}, if for any $\epsilon, \delta >0$ and large enough file size $F$, there exists an $(\epsilon, \delta)$-private scheme. 
\end{mydef}

The object of interest in this paper is the optimal server transmission rate $R^{\star}_{P}(M)$ for normalized cache memory size $M$, given by
\begin{equation}
\label{Eqn:OptimalRate}
R^{\star}_{P}(M)\triangleq \inf \{R : (M,R)\textrm{ is privately achievable}\}.
\end{equation}

\section{Main Results}
\label{Sec:Results}
The main result of this paper is an approximate characterization of the optimal server transmission rate $R^{\star}_{P}(M)$ for any normalized cache memory size $M$. We propose a private caching and delivery scheme to show the following upper bound on $R^{\star}_P(M)$.
%
\begin{theorem}
For $M = \frac{Nt}{K-t}+1$ with $t\in\{0,1,\ldots,K-2\}$, the following rate is privately achievable
\begin{equation}
\label{Eqn:AchievableRate}
R_C(M) \triangleq \frac{K(N+M-1)}{N+(K+1)(M-1)}.
\end{equation}
For $M = N(K-1)$, we achieve the rate $R_C(M)=1$. Further, for any general $1\leq M \leq N(K-1)$, the convex envelope of these points is achievable.
\label{Thm:AchievableRate}
\end{theorem}
Some comments are in order. Note that the achievable rate $R_C(M)  = 1$ for $M = N(K-1)$. This is in fact the minimum achievable rate for any private caching scheme, i.e. $R^{\star}_{P}(M) \ge 1, \ \forall \ M$. Intuitively, this is because the information leakage as defined in \eqref{sec} is constrained to be negligible for any private scheme, and this implies that the contents of a cache cannot provide any information about the requested file on its own. Hence, for a user to learn the file it requested, it must receive from the server at least $F$~bits. We provide a formal proof in Section~\ref{sec:converse}. Similarly, note that we only consider $M \ge 1$ in the above result. As we prove in Section~\ref{sec:converse}, this is indeed a necessary condition for the existence of a private caching and delivery scheme.

The next result provides an information-theoretic lower bound on the server transmission rate of any private caching and delivery scheme. 
\begin{theorem}
For $1\leq M \leq N(K-1)$,
\begin{equation}
\label{Eqn:LowerBound}
R_P^{\star}(M)\geq \max_{s \in \{1,2,...,\min\{N/2,K\}\}} \frac{s\floor{N/s}-1-(s-1)M}{\floor{N/s}-1}.
\end{equation}
\label{Thm:LowerBound}
\end{theorem}
The above result is obtained using cut-set based arguments and is presented in section~\ref{sec:converse}. The lower bound can be further improved by using non-cut set based arguments as shown in Section~\ref{sec:discussion}. However, the cut-set lower bound is indeed tight for the case with $N=K=2$, as shown in Section~\ref{achievability}. This lower bound also suffices to show that, in general, the server transmission rate of the proposed scheme is within a constant factor of the optimal for most regimes of interest: 
\begin{theorem}
For $M \ge M_0 \triangleq 1+\max\left\{0, \frac{N(K-N)}{(N-1)K+N}\right\}$, 
\begin{equation}
1  \le \frac{ R_C(M)}{R^{\star}_P(M) } \leq 16.
\end{equation}
%
\label{Thm:OrderOptimality}
\end{theorem}
\blu{The above theorem establishes the order-optimality of our scheme. The constant 16 can perhaps be improved using more sophisticated converse arguments, see, for example, \cite{ghasemi2015improved, sengupta2015improved}.}

It is easy to verify that $M_0 = 1$ for $N \ge K$. Recall that $M \ge 1$ is necessary for any private caching and delivery scheme and thus our proposed scheme is order-optimal for all permissible values of the normalised cache memory size $M$. For $N < K$, we have $M_0 \leq 1 + N/(N-1) < 5/2$ and thus the above result establishes the order-optimality of our proposed scheme for all regimes of interest except for $1 \leq M \leq 5/2$. This limitation is because for $N < K$, the lower bound in Theorem~\ref{Thm:LowerBound} depends only on the number of files $N$. However, we expect the optimal rate to increase with the number of users $K$, since we have to ensure privacy for a larger set of users.

\bluer{We further propose a \textit{decentralized} variant of our private caching scheme in Section~\ref{sec:decentralized} and analyze its performance, which results in the following theorem.
\begin{theorem}\label{Thm:Decentralized}
For $M\ge 1$ there exists a decentralized scheme for which the following rate is privately achievable
$$R_D(M)=\begin{cases}
K & 1\le M<2,\\
\frac{(1-(1-q)^K)}{q} & M\ge 2,
\end{cases}$$
where $q \triangleq (M-2)/(M+N - 2)$. Moreover, when compared to the privately achievable rate of the centralized scheme $R_C(M)$ as given in Theorem~\ref{Thm:AchievableRate}, we have 
$$
\frac{ R_D(M) }{R_C(M)} \le 6
$$
for $M\ge 1$ when $N\geq K$ and for $M\ge 5/2$ when $N<K$. 
\end{theorem}
Note that the above result together with Theorem~\ref{Thm:OrderOptimality} implies the order-optimality of the \textit{decentralized} private caching scheme in the above regimes. 
}

\section{Achievability Scheme} \label{achievability}
\label{Sec:Scheme}
\blu{In this section, we first discuss achievability schemes for some example setups, and then present our generalized achievability scheme.}
\subsection{Examples}
\subsubsection{Optimal Scheme for $N=K=2$ and $M=1$}
\begin{figure}
\centering
\includegraphics[scale=.85]{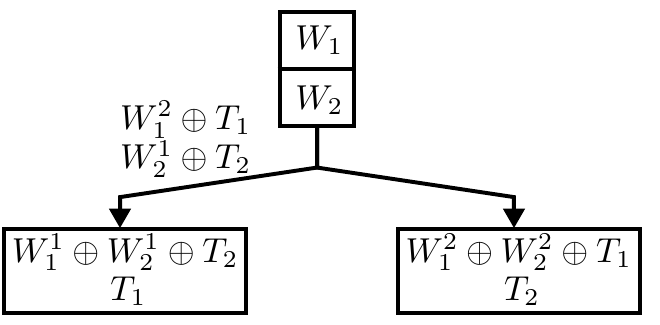}
\caption{Optimal scheme for $N=K=2$, $M = 1$ achieving rate $R_C =1$. Server transmission for the demand vector $(d_1,d_2)=(1,2)$ is shown.}
\label{fig:n2k2a}
\end{figure}
Figure~\ref{fig:n2k2a} shows an example setup with $N = 2$ files and $K = 2$ users with normalized cache memory size $M = 1$. Partition the two files $W_1, W_2$ into two equal parts $W_1^1,W_1^2$ and $W_2^1,W_2^2$ respectively.  
Two independent and uniformly distributed {\em random keys}, $T_1$ and $T_2$ each of size $F/2$ bits, are  generated. During the placement phase, the random keys and their combination with the file parts are put in the caches as shown in Figure~\ref{fig:n2k2a}. During the delivery phase, if the demand vector is $(d_1,d_2)=(1,2)$, the server transmits $W_1^2 \oplus T_1$ and $W_2^1\oplus T_2$, of total size $F$ bits. It can be easily verified that both the users can recover their requested files using their respective cache contents and the server transmission. Furthermore, neither user can derive any information about the file they did not request. Similarly, any other demand vector can also be privately satisfied using a server transmission of size $F$ bits. Specifically, note that when the users demand the same file, the server may send it in the clear. Thus, the memory-rate pair $(M = 1, R= 1)$ is  achievable. As mentioned before, $M \ge 1, R \ge 1$ are necessary conditions for feasibility in our setup and so the scheme presented above is in fact optimal. 

While the scheme described above is optimal, it is not immediately clear how to generalize it to larger number of files and caches. Instead, below we discuss a sub-optimal scheme at two different memory-rate points. This scheme easily generalizes to our order-optimal scheme.
\subsubsection{A Scheme for $M=1$}
\begin{figure}
\centering
\includegraphics[scale=.85]{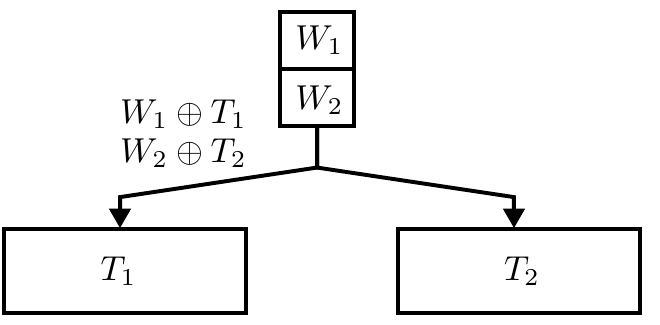}
\caption{Alternate scheme for $N=K=2$, $M=1$ achieving rate $R_C = 2$. Shown is the server transmission for demand vector $(d_1,d_2)=(1,2)$.}
\label{fig:n2k2t0}
\end{figure}
At $M=1$, we cache independent keys $T_i$ of size $F$ bits at each user $i\in [K]$, see Figure~\ref{fig:n2k2t0} for an illustration when $N = K = 2$. During delivery, the server transmits $W_{d_i} \oplus T_i$ for each user $i \in [K]$, resulting in a rate of $K$. It is easy to verify that each user is able to recover its requested file and obtains no information about the other files. Finally, note that the rate of this scheme matches the value of $R_C(1)=K$ (corresponding to $M=1$) in Theorem~\ref{Thm:AchievableRate}. 
\subsubsection{A Scheme for $M=N(K-1)$}
\begin{figure}
\centering
\includegraphics[scale=.85]{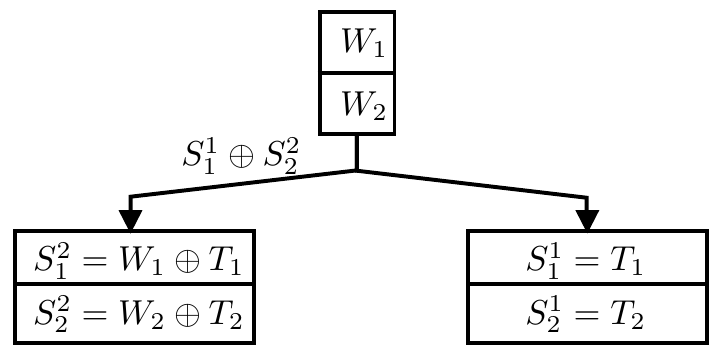}
\caption{Alternate scheme for $N=K=2$, $M=N(K-1)=2$ achieving rate $R_C=1$. Server transmission for the demand vector $(d_1,d_2)=(1,2)$ is shown.}
\label{fig:n2k2t1}
\end{figure}
\remove{
At the other extreme when $M = N(K-1)$, we use a $(K-1,K)$ {\em secret sharing scheme}\footnote{For $m<n$, by an $(m,n)$ secret sharing scheme, we mean a ``scheme'' $p_{S_1,\ldots,S_n|W}$ to generate $n$ equal-sized shares $S_1,\ldots,S_n$ of a uniformly distributed secret $W$ such that any $m$ shares do not reveal any information about the secret and access to {\em all} the $n$ shares completely reveals the secret. i.e., \begin{align*}
I(W;S_A)&=0,\; \forall A\subset [n] \mbox{ s.t. } |A|=m,\\
H(W|S_{[n]})&=0.
\end{align*}%
It is easy to see that, for such a scheme, the shares must have a size of at least $\frac{\log|{\mathcal W}|}{n-m}$ bits, where ${\mathcal W}$ is the alphabet of the secret. When $|{\mathcal W}|$ is large enough, secret sharing schemes which achieve this bound exist~\cite{cramer2015}.}~\cite{shamir1979,cramer2015} for each file $W_i$, $i \in [N]$. For each file $W_i$, $i \in [N]$, the secret sharing scheme provides $K$ shares, each of size $F$ bits and denoted by $\{S_i^{j}\}_{j=1}^{K}$, with the following properties:
}
At the other extreme when $M = N(K-1)$, we use a {\em secret sharing scheme}\blu{~\cite{shamir1979},~\cite{cramer2015}} as defined below\footnote{Note that our terminology is non-standard compared to the literature where an $(t,n)$-{\em threshold} secret sharing scheme has $n$ shares such that any $t$ shares do not reveal any information about the secret and the secret can be reconstructed from any $t+1$ shares.}:
\begin{mydef}
For $m<n$, by an $(m,n)$ {\em secret sharing scheme}, we mean a ``scheme'' $p_{S_1,\ldots,S_n|W}$ to generate $n$ equal-sized shares $S_1,\ldots,S_n$ of a uniformly distributed secret $W$ such that any $m$ shares do not reveal any information about the secret and access to {\em all} the $n$ shares completely reveals the secret. i.e., \begin{align*}
I(W;S_A)&=0,\; \forall A\subset [n] \mbox{ s.t. } |A|=m,\\
H(W|S_{[n]})&=0.
\end{align*}%
\end{mydef}
\blu{For such a scheme, the shares must have a size of at least $\frac{\log|{\mathcal W}|}{n-m}$ bits, where ${\mathcal W}$ is the alphabet of the secret. 
This is easy to see by noting that the first $m$ shares must not reveal any information about the secret, i.e., $I(W;S_1,\ldots,S_m)=0$, and the secret can be completely reconstructed from all the shares taken together, i.e., $I(W;S_1,\ldots,S_n)=H(W)=\log|{\mathcal W}|$. Hence, using the chain rule of mutual information in the second step below,
\begin{align*}
\log|{\mathcal W}| &= I(W;S_1,\ldots,S_n)\\
  &= I(W;S_1,\ldots,S_m) + I(W;S_{m+1},\ldots,S_n|S_1,\ldots,S_m)\\
  &= I(W;S_{m+1},\ldots,S_n|S_1,\ldots,S_m)\\
  &\leq H(S_{m+1},\ldots,S_n|S_1,\ldots,S_m)\\
  &\leq H(S_{m+1},\ldots,S_n)\\
  &\leq (n-m)\log|{\mathcal S}|,
\end{align*}
where ${\mathcal S}$ is the alphabet of each share and, hence, the size of each share in bits is $\log|{\mathcal S}|$. Above, we used the fact that $I(W;S_1,\ldots,S_m)=0$ in the third step and the fact that conditioning cannot lead to an increase in entropy in the next step. The last step follows from the fact that the uniform distribution maximizes entropy for a given alphabet. Hence, the shares must have a size of at least $\frac{\log|{\mathcal W}|}{n-m}$ bits, where ${\mathcal W}$ is the alphabet of the secret. This bound is in fact tight. When $|{\mathcal W}|$ is large enough, secret sharing schemes which achieve this bound exist~\cite{cramer2015}\footnote{\blu{Alternatively, we can view this as a secure network coding problem~\cite{CaiYeung11}, specifically, a form of wiretap channel of type II~\cite{OzarowWyner84}\cite[Section VI.A]{ElRouayhebetal12}, where there are $n$ equal capacity parallel links from the source to destination and there is a wiretapper who can eavesdrop on any $m$ of these links. The results in~\cite{CaiYeung11} imply that, the secure capacity of this network is ${n-m}$ and codes which achieve this capacity exist for sufficiently large alphabet sizes.}}.}

For each file $W_i$, $i \in [N]$, we use a $(K-1,K)$ secret sharing scheme, which provides $K$ shares, each of size $F$ bits and denoted by $\{S_i^{j}\}_{j=1}^{K}$, with the following properties:
\begin{enumerate}
\item[(i)] No collection of $K-1$ shares reveals any information about the file $W_i$, and
\item[(ii)] the file $W_i$ can be recovered from its $K$ shares $\{S_i^{j}\}_{j=1}^{K}$. 
\end{enumerate}
\remove{
\begin{figure}
\centering
\includegraphics[scale=.9]{n3k3t2}
\caption{Alternate scheme for $N=K=3$, $M=4$ achieving rate $R_C =1$. Server transmission for the demand vector $(d_1,d_2,d_3)=(1,2,3)$ is shown.}
\label{fig:n3k3t2}
\end{figure}
}
During the placement phase, different shares are stored in the various caches as follows: the contents of cache $k\in[K]$ is given by $Z_k = \{S_i^j \ : \ i \in [N], j \in [K], j \ne k\}$. Note that there are $N(K-1)$ shares stored in every cache, each of size $F$ bits, and this agrees with the normalized cache memory size $M = N(K-1)$. Next, during the delivery phase, each user requests a file and the server transmits $\oplus_{k\in [K]}S_{d_k}^{k}$ of size $F$ bits, resulting in a rate of $1$. \remove{See Figure~\ref{fig:n3k3t2} for an illustration when $N = K = 3$.}

Since each user $k\in [K]$ already has all the shares $\{S_{i}^{j} \}_{i \in [N], j\neq k}$, the missing share of the demanded file $S_{d_k}^k$ can be obtained, and the file $W_{d_k}$ can be reconstructed. Furthermore for any other file than the one requested, each user $k\in [K]$ only has $(K-1)$ shares which do not reveal any information because of the properties of the $(K-1,K)$ secret sharing scheme. Again, note that the rate of the proposed scheme agrees with the value of $R_C(M =N(K-1))=1$ in Theorem~\ref{Thm:AchievableRate}. See Figure~\ref{fig:n2k2t1} for an illustration when $N = K= 2$. Here the $K = 2$ shares for each file $W_i$ are given by $S_{i}^1 = W_i \oplus T_i$ and $S_i^2 = T_i$, where $T_i$ is a random key of size $F$ bits.
\subsection{Generalized Achievability Scheme}
\blu{We now generalize the ideas presented above to obtain a private caching and delivery scheme for all problem parameters $N, K$, and $M$, and characterize its rate to complete the proof of Theorem~\ref{Thm:AchievableRate}.} In fact, we will propose an $(\epsilon = 0, \delta = 0)$-private scheme, i.e. the probability of error as well as the information leakage are both zero. 

We have already discussed the schemes which achieve $R_{C}(M)$ as defined in Theorem~\ref{Thm:AchievableRate} at $M = 1$ and $M = N(K-1)$. Next, we consider $M = Nt/(K-t) + 1$ for some $t\in\{1,...,K-2\}$. We use a $({K-1 \choose t-1},{K \choose t})$ secret sharing scheme to create ${K \choose t}$ shares, each of size $F_s = \frac{F}{{K \choose t}-{K-1 \choose t-1}}= \frac {Ft}{(K-t){K-1 \choose t-1}}$ bits, for each file $W_i$, $i\in [N]$.
For each file $W_i$, we denote its shares by ${\D}_i \triangleq \{S_i^{\La} \ : \ \La \subset [K], |\La| = t \}$ and define $\C_i^k \triangleq \{ S_i^{\La}: \La \subset [K], |\La| = t, k \in \La\}$. Then for any $k \in[K]$, the shares satisfy the following properties~\footnote{\blu{We note that the properties detailed in \eqref{secsha1}, \eqref{secsha2}, and \eqref{secsha3} do not require each file $W_i$  to be uniformly distributed over $[2^F]$, as long as they are all independent.}}
\begin{align}
\label{secsha1} & I(W_{[N]};\bigcup_{i\in [N]}\C_i^k) = 0 ,\\ 
\label{secsha2} & I(W_{[N]\setminus \{d_k\}}; \bigcup_{i\in [N]}\C_i^k\cup\D_{d_k}) = 0 ,\\
\label{secsha3} & H(W_{d_k}| \D_{d_k}) = 0 .  
\end{align}
The identities \eqref{secsha1}, \eqref{secsha2} imply that ${K-1 \choose t-1}$ shares of a file reveal no information about it and shares of one file do not provide information about another file since they are independent; and \eqref{secsha3} implies that ${K \choose t}$ shares of a file are sufficient for recovering it without error. 
\begin{figure}
\centering
\includegraphics[scale=.9]{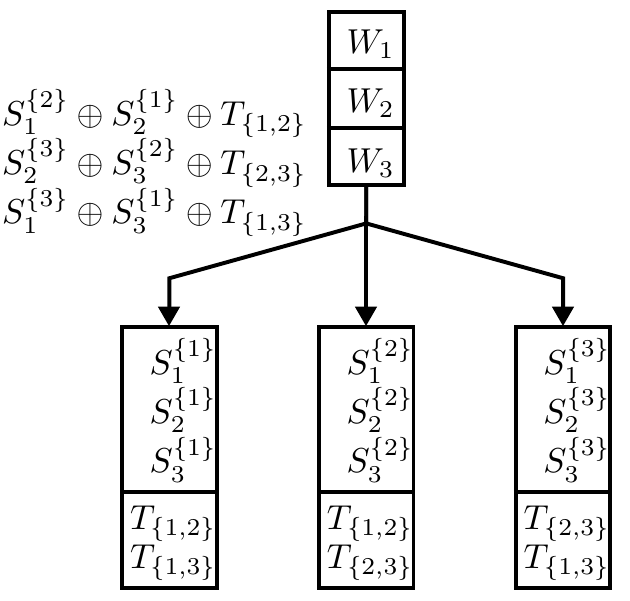}
\caption{General scheme for $N=K=3$ with $t=1, M = Nt / (K - t) + 1 = 5/2$ achieving rate $R_C=3/2$. Server transmission for the demand vector $(d_1,d_2,d_3) = (1,2,3)$ is shown.}
\label{fig:n3k3t1}
\end{figure}
During the placement phase, share $S_i^{\La}$ is placed in the cache of user $k$ if $k \in \La$. Thus $\bigcup_{i\in [N]}\C_i^k$ precisely denotes the shares cached at user $k$. Since we have ${K-1 \choose t-1}$ shares of each of the $N$ files in every user cache, the total memory size in bits needed for storing the shares is given by
\begin{equation}
\label{Eqn:Shares}
F_s \cdot N \cdot {K-1 \choose t-1} =  \frac {Ft}{(K-t){K-1 \choose t-1}}  \cdot N \cdot {K-1 \choose t-1} = \frac{Nt \cdot F}{K-t} .
\end{equation}
In addition to the shares, for each subset $\V \subset [K]$ of users of size $|\V |=t+1$, an independently and uniformly generated key $T_{\V}$ of size $F_s$ bits \blu{(indexed by the subset $V$)} is cached at each user $k \in \V$. \blu{Since each user is part of exactly $K-1 \choose t$ such subsets of size $t+1$, the} cache memory in bits needed to store the keys is given by 
\begin{equation}
\label{Eqn:Keys}
F_s{K-1 \choose t}= \frac {Ft}{(K-t){K-1 \choose t-1}} \cdot {K-1 \choose t} = F .
\end{equation}
Combining \eqref{Eqn:Shares} and \eqref{Eqn:Keys}, the total memory needed per cache is given by $(\frac{Nt}{K-t} + 1)F$ bits which agrees with $M = Nt / (K-t) + 1$. See Figure~\ref{fig:n3k3t1} for an illustration of the placement phase when $N = K = 3$ and $t = 1$. 

During the delivery phase, the demand vector $(d_1,...,d_K)$ is revealed to the server and the users. Then for each $\V \subset  [K]$ such that $|\V |=t+1$, the server transmits $T_{\V } \oplus_{k \in \V } S_{d_k}^{\V \setminus \{k\}}$ on the shared link to the users. See Figure~\ref{fig:n3k3t1} for an example. Consider one such subset $V$ and its associated server transmission. From the placement phase, each $k \in V$ has the key $T_V$ as well as all the shares in the message except $S_{d_k}^{\V \setminus \{k\}}$, and hence each user $k$ can recover the share $S_{d_k}^{\V \setminus \{k\}}$. It is easy to verify that at the end of the delivery phase, each user $k$ would possess all the ${K \choose t}$ shares of its requested file $W_{d_k}$ and thus, from \eqref{secsha3}, can recover it without error. Furthermore, the scheme ensures that the server transmissions do not reveal any information to a user about files it did not request. This combined with \eqref{secsha1}, \eqref{secsha2} ensures that the information leakage, as defined in \eqref{sec}, of the placement and delivery phases of the proposed scheme is zero. Thus, we have a private caching and delivery scheme. Finally, the server transmission size in bits of our proposed scheme at $M = Nt / (K-t) + 1$ is given by 
$$
{K \choose t+1} \cdot  F_s = \frac{ {K \choose t} t \cdot F}{(K-t) {K-1 \choose t-1}} = \frac{KF}{1 + t}.
$$
Substituting $t = (M -1 )K / (N + M - 1)$, we obtain the achievable rate expression $R_C(M)$ as defined in Theorem~\ref{Thm:AchievableRate}.

\section{Lower Bound and Order-Optimality} \label{sec:converse}
In this section, we provide a lower bound on the optimal server transmission rate $R_{P}^{\star}(M)$, as defined in \eqref{Eqn:OptimalRate}. Our proof follows along similar lines as \cite{CachingUM, sengupta2015}, adapted suitably to further take into account the privacy constraint at the users.

Consider a tuple $(M, R)$ which is privately achievable. Fix $s \in [\min\{N/2,K\}]$ and consider users $1,2 \ldots, s$. Suppose user $i\in [s]$ requests file $i$. Since the tuple $(M, R)$ is privately achievable, for any $\epsilon > 0$, there exists a private placement and delivery scheme such that each user $i$ can recover its requested file with a server transmission of rate $R$ along with its cache content $Z_i$ with probability of error at most $\epsilon$. Furthermore for any $\delta > 0$, the information leakage to any user about a file other than the one it had requested is at most $\delta$. Next, consider another scenario where each user $i \in [s]$ requests file $s + i$. Again, since the tuple $(R, M)$ is privately achievable, the recovery and privacy conditions still hold true. 

\begin{figure}
\centering
\includegraphics[scale=1]{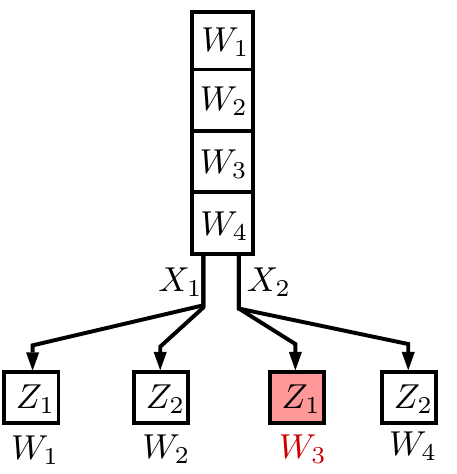}
\caption{\blu{Illustration of the notation for the lower bound argument for $N=K=4$ and corresponding to $s=2$, $k=1$ and $l=2$ is shown. The demand vectors in the two request instances are given by $(d_1^1 = 1, d_2^1 = 2)$ and $(d_1^2 = 3, d_2^2 =4)$ respectively. Here $\widehat{\W}=(W_1,W_2,W_4)$, $\widehat{\Z}=Z_2$, $\widehat{\X}=X_1$, and the resulting bound is $R^{\star}_{P}(M)\ge 3-M$.}}
\label{fig:lb}
\end{figure}

One can repeat the same argument for $\lfloor N/s\rfloor$ different request patterns. Let $X_l$ denote the server transmission corresponding to the $l^{\mbox{\scriptsize th}}$ request instance when the demand pattern is given by $(d_1^l = (l-1)s + 1, d_2^l = (l-1)s + 2, \ldots, \blu{d_s^l} = ls)$. Define $\X_{[\lfloor N/s\rfloor]} \triangleq (X_i : i \in [\lfloor N/s\rfloor])$ and recall that $\Z_{[s]} = (Z_i: i\in [s]), \W_{[s\lfloor N/s\rfloor]} = (W_i: i\in [ s\lfloor N/s\rfloor] )$. Also, given some $k \in [K]$,  let $\widehat{\W} \triangleq \W_{ [s\floor{N/s}] \setminus \{d_k^l\} } = \W_{ [s\floor{N/s}] \setminus \{(l-1)s+k\} }$, $\widehat{\X} \triangleq \X_{[\lfloor N/s\rfloor] \setminus \{l\}}$ and  $\widehat{\Z} \triangleq \Z_{[s]\setminus \{k\}}$. In words, $\widehat{\W}$ denotes the vector of all demanded files except the one requested by the user $k$ in the $l^{\mbox{\scriptsize th}}$ request instance, $\widehat{\X}$ denotes the vector of server transmissions in all the request instances except the $l^{\mbox{\scriptsize th}}$ one, and $\widehat{\Z}$ refers to the contents of all the user caches except the  $k^{\mbox{\scriptsize th}}$ one. \blu{This construction is illustrated in Figure~\ref{fig:lb}.}

Then, from the recovery and privacy conditions, for $(M, R)$ to be privately achievable,\remove{{\color{red} for $\gamma = f(\epsilon, \delta) > 0$}} for $l \in [\lfloor N/s\rfloor]$, $k \in [s]$, we have using Fano's inequality in~\eqref{err} and using~\eqref{sec}	
\begin{align}
\label{lbrec} & H(\W_{[s\floor{N/s}]}|\X_{[\lfloor N/s\rfloor]},\Z_{[s]}) \leq H_b(\epsilon) + \epsilon NF \\
\label{lbsec} & I(\widehat{\W}; X_l, Z_k)\leq \delta 
\end{align}
where for any $x \in [0,1]$, $H_b(x)$ is the binary entropy function. Then, we have
\begin{align*}
(s\floor{N/s}-1)F &= H(\widehat{\W}) \\
&= I(\widehat{\W};\X_{[\lfloor N/s\rfloor]},\Z_{[s]}) + H(\widehat{\W}\ | \ \X_{[\lfloor N/s\rfloor]},\Z_{[s]}) \\
&\overset{(a)}{\leq} I(\widehat{\W};\X_{[\lfloor N/s\rfloor]},\Z_{[s]}) + H_b(\epsilon) + \epsilon NF \\
&= I(\widehat{\W};X_l,Z_k) + I(\widehat{\W};\widehat{\X},\widehat{\Z} \ | \ X_l,Z_k) + H_b(\epsilon) + \epsilon NF \\
&\overset{(b)}{\leq}  I(\widehat{\W};\widehat{\X},\widehat{\Z}| X_l,Z_k) + H_b(\epsilon) + \epsilon NF + \delta \\
&\leq H(\widehat{\X},\widehat{\Z}) + H_b(\epsilon) + \epsilon NF  + \delta\\
&\leq \sum_{i=1, i\neq l}^{\floor{N/s}} H(X_i) + \sum_{j=1, j\neq k}^{s} H(Z_j) + H_b(\epsilon) + \epsilon NF + \delta \\
&\leq (\floor{N/s}-1)R^{\star}_{P}(M)F+ (s-1)MF + H_b(\epsilon) + \epsilon NF + \delta
\end{align*}
where $(a), (b)$ follow from \eqref{lbrec}, \eqref{lbsec} respectively. Rearranging the terms, we get 
$$ 
R^{\star}_{P}(M) \ge \frac{s\floor{N/s}-1-(s-1)M- (H_b(\epsilon) + \epsilon NF  + \delta) / F}{\floor{N/s}-1} . 
$$
The statement of Theorem~\ref{Thm:LowerBound} then follows by noting that the above inequality holds true for any $s \in \{1,2,...,\min\{N/2, K\}\}$ and by choosing $\epsilon,\delta$ to be arbitrarily small.

We now show that the achievable rate $R_C(M)$ from Theorem~\ref{Thm:AchievableRate} is within a constant factor with the above information theoretic lower bound. \blu{In particular, we prove Theorem~\ref{Thm:OrderOptimality}, the details of which may be found in Appendix~\ref{AppB}. Theorem~\ref{Thm:OrderOptimality} states that for $M\geq 1+\max\{\frac{N(K-N)}{(N-1)K+N},0\}$, $$\frac{R_C(M)}{R^{\star}_P(M)}\leq 16.$$}
\blu{In this context, with regards to the condition on $M$, first note that $M\ge 1$ is necessary for privacy. For $N < K$, the additional memory requirement of $\frac{N(K-N)}{(N-1)K+N}$ appears since our proof of order-optimality for this case works when the achievable rate $R_C \le \min\{N, K\} = N$. From Theorem~\ref{Thm:AchievableRate}, it is easy to verify that this requirement is satisfied only when $M\geq 1+ \frac{N(K-N)}{(N-1)K+N}$. In particular, $R_C(M=1)=K > N$.}
\label{Sec:Converse}

\section{Decentralized Scheme} \label{sec:decentralized}
\label{Sec:Dec}
The $\textit{private}$ coded caching scheme discussed in Section~\ref{achievability} requires prior knowledge of the number of users $K$, during the placement phase. This is a disadvantage and our proposed scheme becomes inapplicable if users choose to leave or seek to join the system before the delivery phase. With this motivation, we briefly discuss a {\em randomized}, \textit{decentralized} variant of our earlier \textit{centralized} scheme, wherein the placement phase is carried out at each user independently of the other users. The scheme is inspired by the ideas presented in \cite{DCachingUM}. \remove{{However, unlike~\cite{DCachingUM}, which has zero probability of error, our scheme fails with a probability of error which vanishes as the file size $F$ goes to $\infty$. Note that our definition of {\em privately achievable} (Definition~\ref{def:privately-achievable}) allows for such a vanishing probability of error.}} We further show that this scheme, while being decentralized, is still order-optimal with respect to the information-theoretic lower bound.

As mentioned before, \cite{sengupta2015} studied the problem of secure coded caching, where the goal is to protect information about the files from an adversary who eavesdrops on the server transmissions. The authors of this paper also presented a decentralized scheme for this problem. However, the scheme presented there employed a centralized key placement phase which required the knowledge of the total number of users and the content cached at each user in the placement phase. In contrast, the scheme we propose has a decentralized key placement phase as well. \bluer{Employing such a decentralized key placement entails an additive penalty in terms of the expected rate, which is shown to vanish as the file size $F$ goes to $\infty$.

We now present the details of our decentralized scheme. In the placement phase, a {`private'} key $C_i$ of size $F$ bits is placed at each user $i\in [K]$. This consumes $F$ bits at each user. Additionally, if $M\ge 2$, for each file $W_i, i\in[n]$, we generate $G \in \mathbb{N}$ shares, denoted by $\D_i$. These shares are such that the file $W_i$ can be reconstructed using $\D_i$, but any subset of $qG$ shares\footnote{We ignore the integrality constraint in this section for ease of exposition. We can approximate the rates shown as closely as desired by choosing a large enough file size $F$.} from $\D_i$ provide no information about the file $W_i$, where 
{\[q \triangleq (M-2)/(M+N+r - 2).\] and $r$ is a small positive constant.} Let $h$ denote the size of each share. By working with a large enough $F$, we can obtain a share size of (approximately) $h=F/(G-qG)$ bits and thus
\begin{equation*} \label{eqndecsize}
Gh = F/(1-q).
\end{equation*}
Then for each user $k \in [K]$ and for each file $W_i, i\in[n]$, we independently pick a subset of $qG$ shares from $\D_i$ uniformly at random and cache them at user $k$. Next, we generate a collection of {$G\cdot((1-q)/q+r)$} independent keys $\U$, where each key is of size $h$ bits. Then for each user $k \in [K]$, we independently pick a subset of {$((1-q)+rq)\cdot G$} keys from $\U$ uniformly at random and cache them at user $k$. We refer to these keys as {`shared'} keys. Thus, after placement, if $M\ge 2$, the total memory used at each cache in bits is given by {
\begin{align*}
qG\cdot h \cdot N+((1-q)+rq)G \cdot h + F& = \left(q(N+r) + 1 - q\right) \cdot Gh +F \\
& = \left(q(N+r) + 1 - q \right) \cdot F / (1 - q) +F\\
&= \left( q(N+r) / (1-q) + 2 \right) \cdot F\\
&= \left( \frac{(M-2)(N+r) /(M+N+r - 2)}{(N+r) / (M + N+r - 2) } + 2 \right) \cdot F\\
&= MF.
\end{align*}}
If $1 < M < 2$, the memory at each cache is used only for storing a private key, which occupies $F$ bits. Thus in either case, the memory constraint is satisfied at each cache. 

For any subset of users $S \subseteq [K]$ and any file $W_i$, let $\G_i^S$ denote the shares of file $W_i$ that are cached\footnote{$\G_i^\phi$ refers to those shares which are not cached at any user.} exclusively at all the users in this subset $S$, i.e., these shares are cached at each user in $S$ and not cached at any user in $[K]\setminus S$. It is easy to verify that $\mathbb{E}[|\G_i^S|] = q^{|S|}(1-q)^{(K-|S|)}G$. Similarly, let $\U^S$ denote the set of shared keys cached exclusively in the subset $S$, then {$\mathbb{E}[|\U^S|] = q^{|S|}(1-q)^{(K-|S|)}\left( (1-q)/q + r \right)G$. Here we also note that 
\begin{equation}
\label{Dec:Exp1}
\mathbb{E}[|\U^S|-|\G_{d_k}^{S \setminus k}|]= q^{|S|}(1-q)^{(K-|S|)}rG. 
\end{equation}}
During the delivery phase, the users reveal the demand vector $(d_1,d_2,...,d_K)$ to the server. If $1 \le M \le 2$, the server simply transmits $K$ vector sums, $W_{d_k}\oplus C_k$, $k\in[K]$. Recall that $C_i$'s are the private keys stored in each cache. In this case, the rate $R_D(M)=K$. 

In the analysis that follows, we assume $M>2$. {Let $\Q$ denote the event that  the following condition is satisfied
\begin{align}
|\U^S| \ge \max_{k \in S}|\G_{d_k}^{S \setminus k}| \label{eq:distributedkeysize},
\end{align}
for each non-empty subset $S \subseteq [K]$. If $\Q$ occurs, then for every non-empty subset $S \subseteq [K]$, the server transmits a vector sum\footnote{We zero pad the vectors $\G_{d_k}^{S \setminus k}$ before summing so that they are all of the same size, namely, $|\U^S|$.} of file shares encrypted with the corresponding shared key as $\left(\oplus_{k \in S} \G_{d_k}^{S \setminus k}\right) \oplus \U^S $. For each subset $S$, this allows every user $k \in S$ to recover $\G_{d_k}^{S \setminus k}$ since it has access to $\U^S$ and $\G_{d_j}^{S \setminus j}$ for all $j \in S \setminus k$. Thus, from the various server transmissions, each user $k$ gains access to all the shares of its requested file  $W_{d_k}$ which enables its reconstruction. If $\Q$ does not occur, the server transmits $K$ vector sums, $W_{d_k}\oplus C_k$, $k\in[K]$. Since each user $k\in [K]$ already has the private key $C_k$, it can recover the requested file $W_{d_k}$.

We now argue that our privacy condition (\ref{sec}) is satisfied by the proposed scheme. The placement phase provides a user with no more than $qG$ shares of any file, so no information about the files is leaked during this phase. First, consider the case when the condition for event $\Q$ (\ref{eq:distributedkeysize}) is satisfied. {Recall that in this case, for each non-empty subset $S \subseteq [K]$, the server transmits an encrypted sum of shares $\left(\oplus_{k \in S} \G_{d_k}^{S \setminus k}\right) \oplus \U^S$. The occurrence of event $\Q$ implies that for each subset $S\subseteq[K]$, the sizes of the shared keys satisfy $|\U^S| \ge \max_{k \in S}|\G_{d_k}^{S \setminus k}|$. Thus, each user obtains access only to the shares of its requested file $W_{d_k}$ in the delivery phase, and no new information about the shares of other files is revealed. Next, consider the case when $\Q$ does not occur. In this case, the server transmits $W_{d_k}\oplus C_k$ for each $k \in [K]$. Since the private key $C_k$ is present only at user $k$, our privacy condition (\ref{sec}) holds.}

{We now show a lower bound on $\mathbb{P}[\Q]$. For this, define $p\triangleq q^{|S|}(1-q)^{(K-|S|)}$ so that (\ref{Dec:Exp1}) becomes
\begin{equation}
\label{Dec:Exp}
\mathbb{E}[|\U^S|-|\G_{d_k}^{S \setminus k}|]= p\cdot rG.
\end{equation}
The following can also be easily shown
\begin{align}
\label{Dec:VarG} \text{Var}[|\G_{d_k}^{S \setminus k}|]\le p(1-q)/q \cdot G \ , \\
\label{Dec:VarU} \text{Var}[|\U^S|] \le p\left( (1-q)/q + r \right)G \ .
\end{align}
}
Then, we have 
\begin{align*}
\mathbb{P}(|\U^S|<|\G_{d_k}^{S\setminus k}|)&= \mathbb{P}(|\U^S|-|\G_{d_k}^{S\setminus k}|-p\cdot rG<-p\cdot rG) \\
&\le \mathbb{P}(||\U^S|-|\G_{d_k}^{S\setminus k}|-p\cdot rG|>p\cdot rG) \\
&\overset{(a)}{\leq} \frac{\text{Var}[|\U^S|-|\G_{d_k}^{S\setminus k}|]}{p^2r^2G^2} \\
&\overset{(b)}{=} \frac{\left(\text{Var}[|\U^S|]+\text{Var}[|\G_{d_k}^{S\setminus k}|]\right)}{p^2r^2G^2} \\
&\overset{(c)}{\leq} \frac{\left( p((1-q)/q + r)G + p(1-q)/q\cdot G \right)}{p^2r^2G^2} \\
&= \frac{\left( 2(1-q)/q+ r \right)}{pr^2G}\\
&= \frac{\left( 2(1-q)/q+ r \right)}{q^{|S|}(1-q)^{K-|S|}r^2G},
\end{align*}
where $(a)$ follows from (\ref{Dec:Exp}) and Chebyshev's inequality; and $(b)$ follows from the fact that $|\U^S|$ and $|\G_{d_k}^{S\setminus k}|$ are independent. Finally, $(c)$ is obtained by plugging in~(\ref{Dec:VarG}),~(\ref{Dec:VarU}). Applying the union bound over all $k \in S$ and $S\subseteq [K]$, we thus have 
\begin{align}
\nonumber
\mathbb{P}[\Q]&\ge 1-\sum_{j=1}^{K}{K\choose j}j\cdot \frac{\left( 2(1-q)/q+ r \right)}{q^j(1-q)^{K-j}r^2G}\\
\label{qbound}&= 1-\frac{K(2(1-q)/q+r)}{q^K(1-q)^{K-1}r^2G}.
\end{align}
Now the expected server transmission rate $\mathbb{E}[R_D(M)]$ can be bounded by first considering whether or not $\Q$ occurs and then by going over all non-empty subsets $S \subseteq [K]$. It is given by
\begin{align*}
\mathbb{E}[R_D(M)]F &= \mathbb{P}[\Q]\sum_{S\subseteq K, S\neq \phi}\mathbb{E}[(|\U^S|)|\Q]h+(1-\mathbb{P}[\Q])KF\\
&\overset{(a)}{\leq} \sum_{S\subseteq K, S\neq \phi}\mathbb{E}[|\U^S|]h+(1-\mathbb{P}[\Q])KF\\
&\overset{(b)}{=} \sum_{j=1}^{K}{K \choose j}q^{j}(1-q)^{K-j}\left((1-q)/q+r\right)Gh +(1-\mathbb{P}[\Q])KF\\
&\overset{(c)}{\le} Gh(1-(1-q)^K)\left((1-q)/q+r\right)  + \frac{K(2(1-q)/q+r)}{q^K(1-q)^{K-1}r^2G}\cdot KF\\
&= F\left((1-(1-q)^K)\left(\frac{1}{q}+\frac{r}{1-q}\right)  + \frac{K^2(2(1-q)/q+r)}{q^K(1-q)^{K-1}r^2G}\right)
\end{align*}
where $(a)$ follows since for any random variable $\Xin\ge 0$ and event $\Y$, $\mathbb{E}[\Xin]=\mathbb{P}[\Y]\mathbb{E}[\Xin|\Y]+\mathbb{P}[\overline{\Y}]\mathbb{E}[\Xin|\overline{\Y}]\ge\mathbb{P}[\Y]\mathbb{E}[\Xin|\Y]$, $(b)$ follows since $\mathbb{E}[|\U^S|]=  q^{|S|}(1-q)^{(K-|S|)}((1-q)/q+r)G$ and $(c)$ follows from (\ref{qbound}). Since $r$ can be chosen to be any small positive constant, by letting the file size $F$ and the number of shares $G$ grow large, $\mathbb{E}[R_D(M)]$ can be made arbitrarily close to
\begin{equation}
\mathbb{E}[R_D(M)]=\frac{(1-(1-q')^K)}{q'},\label{Eqn:ExpectedRateD}
\end{equation}
where $q'=(M-2)/(M+N-2)$. Further, by the law of large numbers, the server transmission rate $R_D(M)$ approaches $\mathbb{E}[R_D(M)]$ (almost surely) as file size goes to $\infty$. The requirement of large file size for this convergence is similar in spirit to the other works on decentralized caching~\cite{DCachingUM},~\cite{sengupta2015}.} This completes the description of the decentralized version of our proposed private coded caching scheme.

In Figure~\ref{fig:plot_comparison_1}, $R_D(M)$ is plotted against the centralized rate, $R_C(M)$ in Theorem~\ref{Thm:AchievableRate} and the cut-set lower bound in Theorem~\ref{Thm:LowerBound}. In the plot we see that the decentralized rate is very `close' to the centralized rate for a large range of memory sizes. We now quantify this gap.

We first show that for $M\ge 2$, $$R_D(M)/R_C(M-1)\le 2.$$
Note that for $M\ge 2$, $R_C(M-1)$ can be re-written as $R_C(M-1)= 1/\left(q'+\frac{1}{K}\right)$. Thus, we have $R_D(M)/R_C(M-1)=\left(1+\frac{1}{Kq'}\right)(1-(1-q')^K)$. Then, consider the following cases:\\
{\em Case 1: $Kq' \ge 1$.} In this case $$\left(1+\frac{1}{Kq'}\right)(1-(1-q')^K) \leq 1+\frac{1}{Kq'} \leq 2,$$
{\em Case 2: $Kq' < 1$}. Here $$\left(1+\frac{1}{Kq'}\right)(1-(1-q')^K) \leq \left(1+\frac{1}{Kq'}\right)Kq' = Kq'+1 \leq 2$$
and this proves our result.

Moreover it can be easily verified that for $N\geq K$, $M\ge 2$ and for $N< K$, $M\ge 5/2$, $$\frac{R_C(M-1)}{R_C(M)}\leq\frac{N+(K+1)(M-1)}{N+(K+1)(M-2)}\leq 3.$$ Using this and the fact that for $1\leq M \leq 2$, $R_D(M)$ is constant, we have for $N\geq K$, $M\ge 1$ and $N<K$, $M\ge 5/2$, $$R_D(M)/R_C(M) \leq 6.$$
Since in both the above regimes, $N\geq K$, $M\ge 1$ and $N<K$, $M\ge 5/2$, $R_C(M)/R^{\star}_P(M)\leq 16$ from Theorem~\ref{Thm:OrderOptimality}, order-optimality of our decentralized coded caching scheme is thus proven.
\begin{figure}
\centering
\includegraphics[scale=.5]{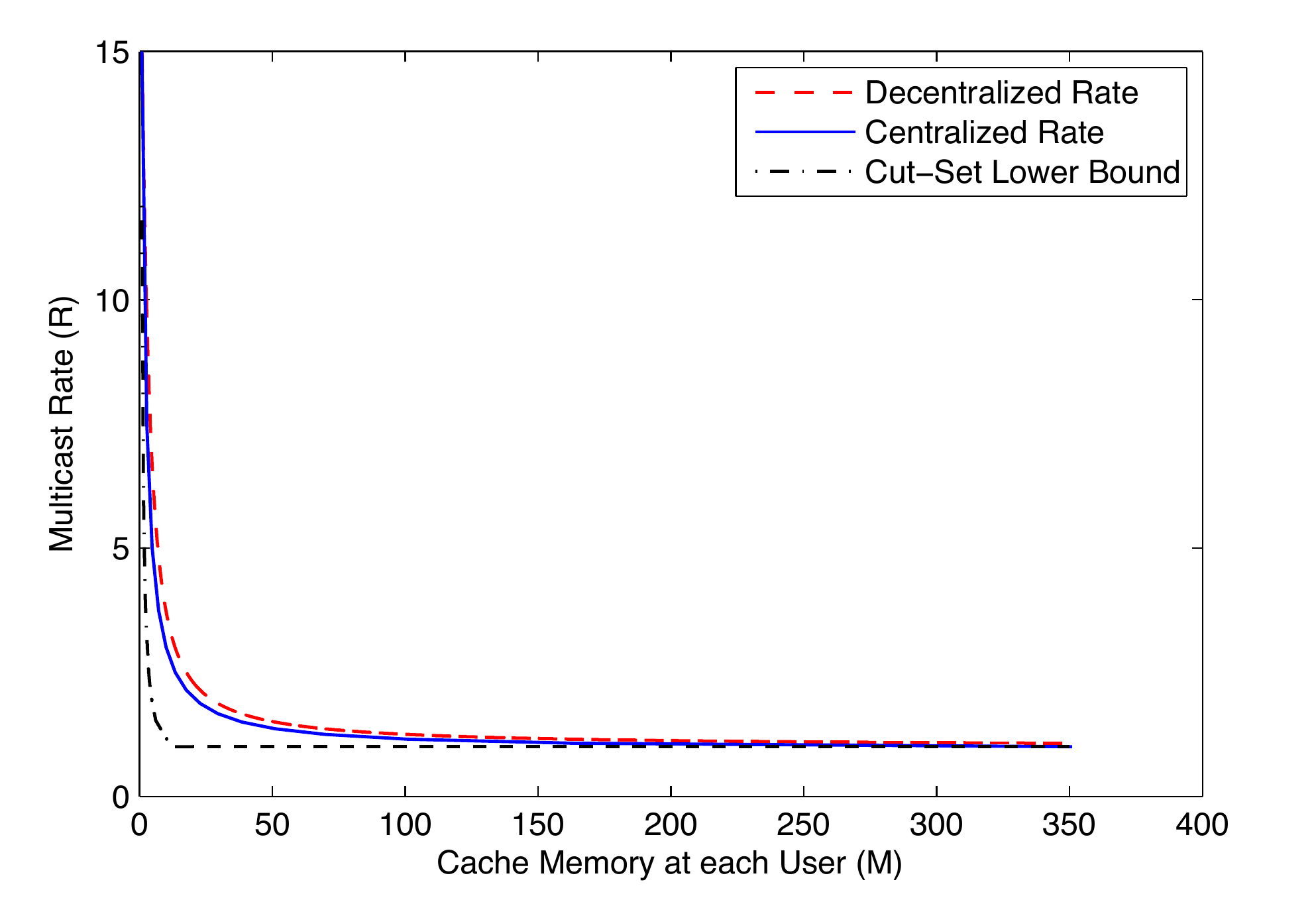}
\caption{The plot shows the achievable rates and the converse for a setup with $N=25$ files and $K=15$ users. The dashed red line is the decentralized rate, $R_D(M)$. The solid blue line is the rate obtained with the earlier centralized scheme, $R_C(M)$. The dash-dot black line is the cut-set lower bound.}
\label{fig:plot_comparison_1}
\end{figure}
}

\section{Discussion} \label{sec:discussion}
\label{Sec:Discussion}
\begin{figure}
\centering
\includegraphics[scale=.5]{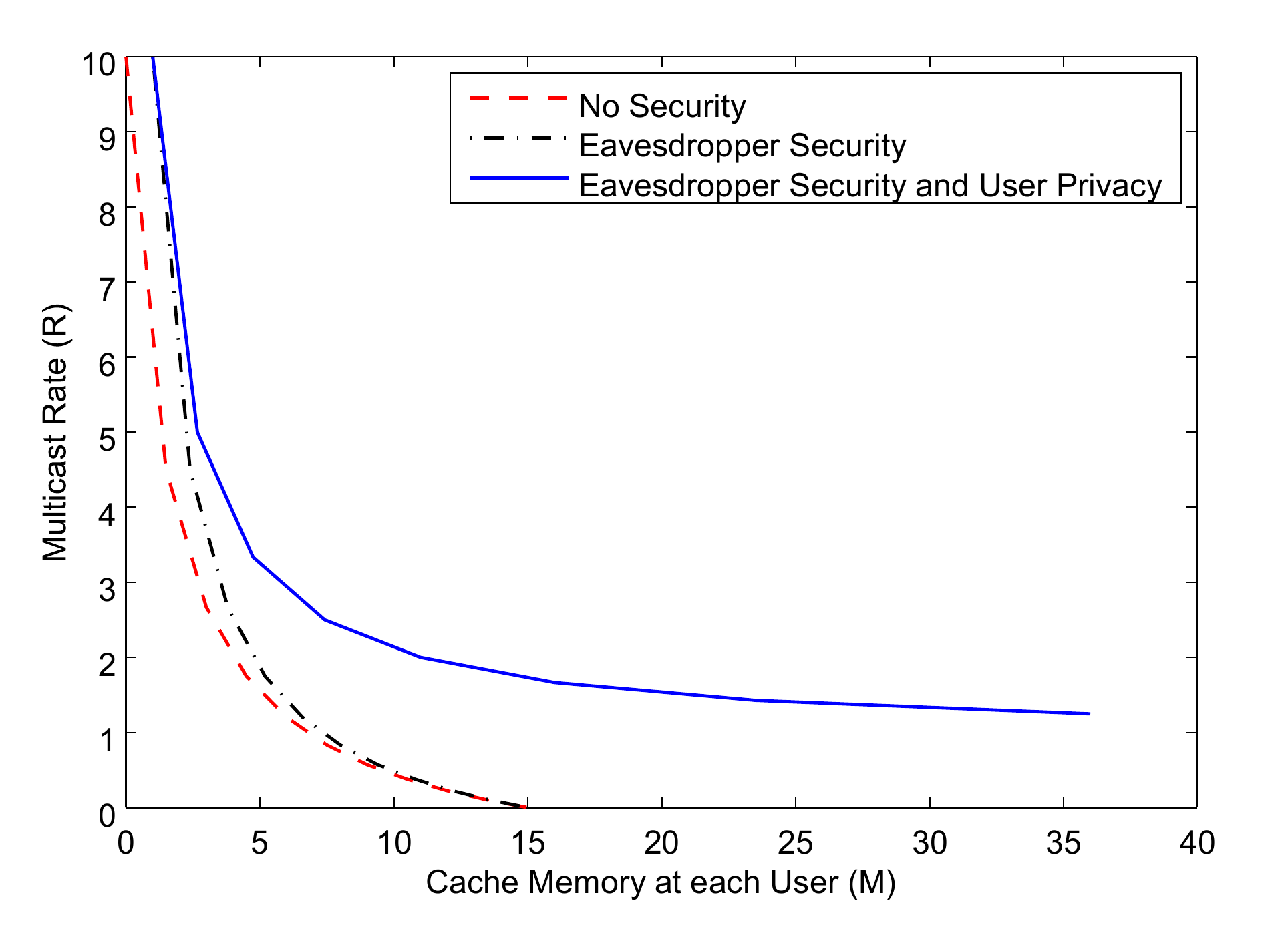}
\caption{The plot shows the achievable rates for a setup with $N=15$ files and $K=10$ users under various conditions. The dashed red line is the achievable rate with no security obtained in \cite{CachingUM}. The dash-dot black line is the achievable rate with only eavesdropper security achieved in \cite{sengupta2015}. The solid blue line is our achievable rate $R_C(M)$ with eavesdropper security and privacy.}
\label{fig:plot_comparison}
\end{figure}
As mentioned before, the work closest to ours is \cite{sengupta2015}, which studied the optimal server transmission rates needed to keep the files secure from an eavesdropper listening to the transmissions on the shared link. In contrast, we imposed the privacy requirement that users should not be able to learn about files they did not request. An obvious scenario of interest is when both the conditions, security against an eavesdropper and privacy against users have to be satisfied.
Let $R_{PE}^{\star}(M)$ denote the optimal server transmission rate in such a setup, as a function of the normalized cache size  $M$. 

As an example, recall the setup in Figure~\ref{fig:n2k2a} with $N=K=2$ and $M=1$ for which the minimum rate for a private scheme is given by $R_P^{\star}(M=1) = 1$. Under the optimal scheme illustrated in Figure~\ref{fig:n2k2a}, when both users request say file $W_1$, the server simply transmits $W_1$ on the shared link. While this sufficed for satisfying the user privacy constraints, clearly it will not work in the presence of an eavesdropper. In fact, the memory-rate tuple $(M=1, R=1)$ is not feasible if we insist on both privacy against users and security against the eavesdropper. The optimal server transmission rate in this scenario is given by $R_{PE}^{\star}(M) = 3 - M$ for $1\le M \le 2$. Proof of this involves certain non-cut set based ideas along the lines of \cite[Appendix]{CachingUM}.

While the optimal scheme in the above example did not protect against eavesdroppers, the general achievability scheme proposed in Section~\ref{Sec:Scheme} does in fact have this additional property since each server transmission to a subset $\V $ of users is protected using a key\footnote{Strictly speaking, this is not true for the scheme at the extreme memory point $M = N(K-1)$ since we do not use a key to protect the server transmission. However, this can be easily fixed without affecting order-optimality by additionally storing a common key in each cache and securing the server transmission with this key.}  $T_{\V }$. Thus, an eavesdropper who has access to these transmissions can obtain no information about the files. This implies that the rate function $R_C(M)$ as defined in \eqref{Eqn:AchievableRate} is in fact achievable for the setup with both security and privacy constraints, i.e. $R_{PE}^{\star}(M) \le R_C(M)$. Furthermore, it is easy to see that the lower bounds in Theorem~\ref{Thm:LowerBound} and the order-optimality result in Theorem~\ref{Thm:OrderOptimality} also continue to hold. Thus, the transmission rate for our proposed scheme is still within a constant factor of the optimal when both security and privacy conditions are imposed. 

Figure~\ref{fig:plot_comparison} plots the order-optimal transmission rates under various constraints. Note that when either no constraint or only the security against eavesdropper constraint is imposed, the achievable rate is zero at $M =N$. On the other hand, once the user privacy condition is activated, the minimum achievable rate for any value of $M$ is one. Furthermore, as the figure illustrates, the gap between the rate with no security and the rate with security against an eavesdropper is not very large. This was in fact shown to be at most a constant factor in \cite{sengupta2015}. The same continues to hold for a large memory regime, $1<M<N\frac{K-1}{2K}$, when a further user privacy constraint is also added.
\remove{
\blu{As mentioned before in Section~\ref{Sec:Dec}, for the case with eavesdropper security alone, a \textit{decentralized} scheme was presented in \cite{sengupta2015}, which we had argued to not be strictly \textit{decentralized}. It turns out that following a approach similar to what we take in Section~\ref{Sec:Dec}, the key placement for the eavesdropper setup in \cite{sengupta2015} can be made strictly \textit{decentralized}.} This may be done by independently caching at each user, a subset of size $q_1.(1-q_1)/q_1.F$ bits chosen uniformly at random from a key stream of size $(1-q_1)/q_1.F$, where $q_1\triangleq (M-1)/(N-1)$. The file placement and delivery procedures are identical to \cite{sengupta2015}. As a result, we obtain the same transmission rate as \cite{sengupta2015}, but via a strictly \textit{decentralized} scheme.}

\begin{appendices}
\section{}\label{AppB}
In this section, we provide details of proof of Theorem~\ref{Thm:OrderOptimality}. We first define $M_S\triangleq M-1$. Then using $N/s-1\leq \floor{N/s}$ and $M_S = M-1$ in Theorem~\ref{Thm:LowerBound}, we have
\begin{align}
\label{eq:rcs} R^{\star}_P(M)\geq \max_{s \in \{1,2,...,\min\{N/2,K\}\}} s-M_S\frac{s(s-1)}{N-2s}.
\end{align}
Also, the achievable rate expression $R_C(M)$ may be written as $$R_C(M)=\frac{K}{1+KM_S / (N+M_S)}.$$
For values of $M$ in our range of interest, if $K\leq N$,
\begin{align*}
& R_C(M)=\frac{K}{1+KM_S/(N+M_S)}\leq K.
\end{align*}
On the other hand, if $K>N$,
\begin{align}\label{mcond}
& R_C(M)\leq R_C\left(\frac{N(K-N)}{(K+1)N-K}\right)=N.
\end{align}
Thus, we have $R_C(M)\leq \min\{N,K\}$.

\noindent
{\em Case 1: $\min\{N,K\}\leq 16$}. \\
In this case, $$R_C(M)\leq \min\{N,K\}\leq 16.$$
And since $R^{\star}_P(M)\geq 1$,
$$\frac{R_C(M)}{R^{\star}_P(M)}\leq 16.$$
{\em Case 2: $\min\{N,K\}> 16$}. \\
In this case we consider 3 regions based on the values of $M_S$. \\
{\em Region I: $0\leq M_S<\max\{N,K\}/(K-1)$. } 

Let $s=\floor{0.205\min\{N,K\}}$ in \eqref{eq:rcs}, using\footnote{$s\ge 1$ holds if $\min\{N,K\} \ge 5$. This assumption is however, not critical for our analysis.} which we obtain
\begin{align}
\nonumber R^{\star}_P(M)& \geq s-M_S\frac{s(s-1)}{N-2s} \\ 
\nonumber &\geq \left. \floor{0.205\min\{N,K\}}-M_S\frac{\floor{0.205\min\{N,K\}}(\floor{0.205\min\{N,K\}}-1)}{N-2\floor{0.205\min\{N,K\}}} \right. \\
\label{reg1} &\geq \min\{N,K\}\left( 0.205-\frac{1}{\min\{N,K\}}-\frac{0.205(0.205\min\{N,K\}-1)(\frac{\max\{N,K\}}{K-1})}{N-2*0.205\min\{N,K\}}\right).
\end{align}
Now consider the expression $\max\{N,K\}(0.205\min\{N,K\}-1)/(K-1)$.
If $N<K$,
\begin{align*}
\max\{N,K\}\frac{0.205\min\{N,K\}-1}{K-1}&=K\frac{0.205N-1}{K-1} \\
&\leq 0.205N(16/15).
\end{align*}
And if $N\geq K$,
\begin{align*}
\max\{N,K\}\frac{0.205\min\{N,K\}-1}{K-1}&=N\frac{0.205K-1}{K-1} \\
&\leq 0.205N \\
&\leq 0.205N(16/15).
\end{align*}
Plugging this into \eqref{reg1} we get 
\begin{align*}
R^{\star}_P(M)&\geq \min\{N,K\}\left( 0.205-\frac{1}{\min\{N,K\}}- \frac{16}{15}\frac{0.205^2}{1-2*0.205\frac{\min\{N,K\}}{N}}\right) \\
& \geq \min\{N,K\} \left( 0.205-\frac{1}{16}-\frac{16}{15}\frac{0.205^2}{1-2*0.205} \right) \\
& \geq \min\{N,K\}/16.
\end{align*}
This gives $$\frac{R_C(M)}{R^{\star}_P(M)}\leq 16.$$
{\em Region II: $\max\{N,K\}/(K-1) \leq M_S<N/15$. }\\
In this region, note that 
\begin{align}
\label{ref1}  &1/M_S \geq (K-1)/\max\{N,K\} 
\end{align}
and that
\begin{align}
\label{ref2} M_S<N/15 \Rightarrow \frac{N}{N+M_S}\geq \frac{15}{16}. 
\end{align} 
Now,
\begin{align*}
R_C(M)&=\frac{K}{1+KM_S/(N+M_S)} \\
&\leq \frac{K}{KM_S/(N+M_S)}\leq \frac{N+M_S}{M_S}.
\end{align*}
Letting $s=\floor{0.198\frac{N+M_S}{M_S}}$ in \eqref{eq:rcs} and following\footnote{Since $M_S<N/15$ in this regime, $s\ge 1$ and $\max\{N,K\}/(K-1) \leq M_S$ guarantees than $s<\min\{N/2,K\}$.} steps similar to the one used to get \eqref{reg1} we have
\begin{align*}
R^{\star}_P(M) &\geq \frac{N+M_S}{M_S} \left( 0.198-\frac{M_S}{N+M_S}-\frac{0.198^2}{N/(N+M_S)-2*0.198/M_S} \right).
\end{align*}
Using \eqref{ref1} and \eqref{ref2} in the above inequality, we get
\begin{align*}
R^{\star}_P(M) &\geq \frac{N+M_S}{M_S}\left( 0.198-\frac{1}{16}-\frac{0.198^2}{15/16-2*0.198}\right) \\
&\geq \frac{N+M_S}{M_S}\frac{1}{16}.
\end{align*}
Hence, $$\frac{R_C(M)}{R^{\star}_P(M)}\leq 16.$$
{\em Region III: $N/15 \leq M_S$.}\\
Note that $N/15 \leq M_S \Rightarrow (N+M_S)/16\leq M_S$, using which
\begin{align*}
& R_C(M)=\frac{K}{1+K\frac{M_S}{N+M_S}}\leq \frac{K}{1+K\frac{1}{16}}\leq 16.
\end{align*}
Using $R^{\star}_P(M)\geq 1$ with the above inequality, we get
$$\frac{R_C(M)}{R^{\star}_P(M)}\leq 16.$$
This proves Theorem~\ref{Thm:OrderOptimality}.
\end{appendices}

\bibliography{journal_abbr,references}

\begin{thebibliography}{10}
\providecommand{\url}[1]{#1}
\csname url@rmstyle\endcsname
\providecommand{\newblock}{\relax}
\providecommand{\bibinfo}[2]{#2}
\providecommand\BIBentrySTDinterwordspacing{\spaceskip=0pt\relax}
\providecommand\BIBentryALTinterwordstretchfactor{4}
\providecommand\BIBentryALTinterwordspacing{\spaceskip=\fontdimen2\font plus
\BIBentryALTinterwordstretchfactor\fontdimen3\font minus
  \fontdimen4\font\relax}
\providecommand\BIBforeignlanguage[2]{{%
\expandafter\ifx\csname l@#1\endcsname\relax
\typeout{** WARNING: IEEEtran.bst: No hyphenation pattern has been}%
\typeout{** loaded for the language `#1'. Using the pattern for}%
\typeout{** the default language instead.}%
\else
\language=\csname l@#1\endcsname
\fi
#2}}

\bibitem{CiscoReport}
``Cisco visual networking index (vni) global mobile data traffic forecast
  update,'' 2013,
  \url{http://www.gsma.com/spectrum/wp-content/uploads/2013/03/Cisco_VNI-global-mobile-data-traffic-forecast-update.pdf}.

\bibitem{Ajaykrishnanetal15}
N.~Ajaykrishnan, N.~S. Prem, V.~M. Prabhakaran, and R.~Vaze, ``Critical
  database size for effective caching,'' in \emph{2015 Twenty First National
  Conference on Communications (NCC)}, Feb 2015.

\bibitem{Borst:2010}
S.~Borst, V.~Gupta, and A.~Walid, ``Distributed caching algorithms for content
  distribution networks,'' in \emph{Proc. IEEE INFOCOM}, Mar. 2010, pp.
  1478--1486.

\bibitem{breslau99}
L.~Breslau, P.~Cao, L.~Fan, G.~Phillips, and S.~Shenker, ``Web caching and
  {Z}ipf-like distributions: Evidence and implications,'' in \emph{Proc. IEEE
  INFOCOM}, Mar. 1999, pp. 126--134.

\bibitem{CaiYeung11}
N.~Cai and R.~W. Yeung, ``Secure network coding on a wiretap network,''
  \emph{IEEE Transactions on Information Theory}, vol.~57, no.~1, pp. 424--435,
  Jan 2011.

\bibitem{cramer2015}
R.~Cramer, I.~B. Damg{\aa}rd, and J.~B. Nielsen, \emph{Secure Multiparty
  Computation and Secret Sharing}.\hskip 1em plus 0.5em minus 0.4em\relax
  Cambridge University Press, 2015.

\bibitem{ghasemi2015improved}
H.~Ghasemi and A.~Ramamoorthy, ``Improved lower bounds for coded caching,''
  \emph{arXiv:1501.06003 [cs.IT]}, 2015.

\bibitem{FemtoCaching}
N.~Golrezaei, K.~Shanmugam, A.~G. Dimakis, A.~F. Molisch, and G.~Caire,
  ``Femtocaching: Wireless video content delivery through distributed caching
  helpers,'' in \emph{Proc. IEEE INFOCOM}, Mar. 2012, pp. 1107--1115.

\bibitem{ji2013wireless}
M.~Ji, G.~Caire, and A.~F. Molisch, ``Wireless device-to-device caching
  networks: Basic principles and system performance,'' \emph{arXiv:1305.5216
  [cs.IT]}, 2013.

\bibitem{CachingUM}
M.~A. Maddah-Ali and U.~Niesen, ``Fundamental limits of caching,'' \emph{IEEE
  Trans. Inf. Theory}, vol.~60, no.~5, pp. 2856--2867, May 2014.

\bibitem{DCachingUM}
------, ``Decentralized coded caching attains order-optimal memory-rate
  tradeoff,'' \emph{IEEE/ACM Trans. Netw.}, pp. 1029--1040, Aug. 2015.

\bibitem{OzarowWyner84}
L.~H. Ozarow and A.~D. Wyner, ``Wire-tap channel ii,'' \emph{AT\&T Bell
  Laboratories Technical Journal}, vol.~63, no.~10, pp. 2135--2157, Dec 1984.

\bibitem{scc2016}
V.~Ravindrakumar, P.~Panda, N.~Karamchandani, and V.~Prabhakaran, ``Fundametal
  limits of secretive coded caching,'' in \emph{Proc. IEEE ISIT}, July 2016,
  pp. 425--429.

\bibitem{ElRouayhebetal12}
S.~E. Rouayheb, E.~Soljanin, and A.~Sprintson, ``Secure network coding for
  wiretap networks of type ii,'' \emph{IEEE Transactions on Information
  Theory}, vol.~58, no.~3, pp. 1361--1371, March 2012.

\bibitem{sengupta2015}
A.~Sengupta, R.~Tandon, and T.~C. Clancy, ``Fundamental limits of caching with
  secure delivery,'' \emph{Information Forensics and Security, IEEE
  Transactions on}, vol.~10, no.~2, pp. 355--370, 2015.

\bibitem{sengupta2015improved}
------, ``Improved approximation of storage-rate tradeoff for caching via new
  outer bounds,'' in \emph{Proc. IEEE ISIT}, June 2015, pp. 1691--1695.

\bibitem{shamir1979}
\BIBentryALTinterwordspacing
A.~Shamir, ``How to share a secret,'' \emph{Commun. ACM}, vol.~22, no.~11, pp.
  612--613, Nov. 1979. [Online]. Available:
  \url{http://doi.acm.org/10.1145/359168.359176}
\BIBentrySTDinterwordspacing

\bibitem{shariatpanahi2015multi}
S.~P. Shariatpanahi, S.~A. Motahari, and B.~H. Khalaj, ``Multi-server coded
  caching,'' \emph{arXiv:1503.00265 [cs.IT]}, 2015.

\bibitem{wang2015fundamental}
S.~Wang, W.~Li, X.~Tian, and H.~Liu, ``Fundamental limits of heterogenous
  cache,'' \emph{arXiv:1504.01123 [cs.IT]}, 2015.

\bibitem{Wessels:2001}
D.~Wessels, \emph{Web Caching}, N.~Torkington, Ed.\hskip 1em plus 0.5em minus
  0.4em\relax O'Reilly, 2001.

\bibitem{zhang2015coded2}
J.~Zhang, X.~Lin, C.-C. Wang, and X.~Wang, ``Coded caching for files with
  distinct file sizes,'' in \emph{Proc. IEEE ISIT}, June 2015, pp. 1686--1690.

\end{thebibliography}
\bibliographystyle{IEEEtrans}

\end{document}